\documentclass[12pt]{article}

\begin{document}
\begin{titlepage}
\begin{center}
\hfill   TUM--HEP--605/05 \\
\hfill   MPP--2005--127 \\

\vskip 1cm


{\large \bf{ LEPTOGENESIS, YUKAWA TEXTURES AND WEAK BASIS INVARIANTS}}

\vskip 1cm

Gustavo C. Branco  \ $^{a,b}$ \footnote{gustavo.branco@cern.ch and 
gbranco@cftp.ist.utl.pt}  
M. N. Rebelo  \ $^{a,c}$
\footnote{margarida.rebelo@cern.ch and
rebelo@ist.utl.pt} 
and 
J. I. Silva-Marcos   \ $^a $  \footnote{Joaquim.Silva-Marcos@cern.ch}  

\vskip 1.0cm
 
$^a$ Departamento de F\'\i sica and Centro de F\' \i sica Te\' orica
de Part\' \i culas (CFTP),
Instituto Superior T\' ecnico, Av. Rovisco Pais, P-1049-001 Lisboa,
Portugal.

\vskip 0.5cm

 $^b$ Physik--Department, Technische Universit\"at M\"unchen
James--Franck--Strasse, D--85748 Garching, Germany

\vskip 0.5cm

$^c$ Max-Planck--Institut f\" ur Physik (Werner--Heisenberg--Institut),\\
D-80805 M\" unchen, Germany 

\end{center}

\vskip 3cm

\begin{abstract}
We show that a large class of sets of leptonic texture zeros
considered in the literature imply the vanishing of certain
CP-odd weak-basis invariants. These invariant conditions
enable one to recognize a flavour model corresponding to a 
set of texture zeros, when written in an arbitrary weak-basis 
where the zeros are not manifest. We also analyse the r\^ ole of
texture zeros in allowing for a connection between leptogenesis 
and low-energy leptonic masses, mixing and CP violation. 
For some of the textures the variables relevant
for leptogenesis can be fully determined in terms of low
energy parameters and heavy neutrino masses.
\end{abstract}

\end{titlepage}

\newpage
\section{Introduction}
The evidence for non-vanishing neutrino masses
provides a clear signal of Physics beyond the Standard Model (SM), 
since in the SM neutrinos are strictly massless.
The simplest extension of the SM which allows for nonvanishing
but naturally small neutrino masses consists of the addition of
right-handed neutrinos, leading to the seesaw mechanism \cite{seesaw}.

In general, the seesaw mechanism framework contains a 
large number of free parameters, in fact many more than
measurable quantities at low energies.
In the literature, there have been various attempts at reducing
the number of seesaw parameters either by introducing
texture zeros and/or by reducing the number of right-handed
neutrinos to two. One could be tempted to
follow a bottom-up approach, using the observed pattern of
lepton masses and mixing to infer about the appropriate
set of texture zeros. Unfortunately this approach is not
feasible, since texture zeros are not 
weak-basis (WB) invariant. This
means that  a given set of texture zeros which arise in a certain
WB may not be  present or may appear in different entries 
in another WB. Indeed, each texture zero ansatz corresponds 
to an infinite set of 
leptonic mass matrices, related to each other by WB transformations.
Needless to say, two sets of leptonic mass matrices related by a 
WB transformation contain the same physics.
This raises a number of questions, such as:  

(i) How can one  recognize a flavour model corresponding to a set of
texture zeros, when written in a different WB,
where the zeros are not explicitly present?

(ii) Do the sets of texture zeros considered in the 
literature imply the vanishing of certain WB invariants?

(iii) Can the physical content of a particular texture
zero anzatz be expressed in terms of relations 
involving WB invariants?

In this paper, we address some of the above questions 
and in particular we show that some of the
sets of texture zeros considered in the literature imply
the vanishing of certain CP-odd invariants.
Conversely, we show that starting from arbitrary leptonic 
mass matrices and imposing the vanishing of certain CP-odd
invariants, together with the assumption of no conspiracy
among the parameters of the Dirac and Majorana neutrino mass
terms, one is automatically led to given sets of texture zeros. 
The relevance of CP-odd invariants in the analysis of texture
zero ans\" atze  
was to be expected, since
texture zeros lead in general to a decrease in the number of 
independent CP-violating phases.

This paper is organized as follows. In the next section, we 
describe the framework and set our notation. In section 3, we 
reexamine the connection \cite{leptogenesis} between leptonic low 
energy physics and leptogenesis \cite{Fukugita:1986hr} , 
in the case of one texture zero and two 
right-handed neutrinos. In section 4 we study the relation between
CP-odd WB invariants and texture zeros. Finally section 5 contains 
our conclusions.

\section{Framework}
Let us consider the above mentioned extension of the
SM which consists of the 
addition of one right-handed neutrino per fermion generation.
After spontaneous gauge symmetry breaking, the following
leptonic mass terms are generated: 
\begin{eqnarray}
{\cal L}_m  &=& -[\overline{{\nu}_{L}^0} m_D \nu_{R}^0 +
\frac{1}{2} \nu_{R}^{0T} C M_R \nu_{R}^0+
\overline{l_L^0} m_l l_R^0] + h. c. = \nonumber \\
&=& - [\frac{1}{2}  n_{L}^{T} C {\cal M}^* n_L +
\overline{l_L^0} m_l l_R^0 ] + h. c.
\label{lep}
\end{eqnarray}
where $M_R$, $m_D$, and $m_l$ denote the  
right-handed neutrino Majorana mass matrix, 
the neutrino Dirac mass matrix and the charged
lepton mass matrix, respectively, with
$n_L = ({\nu}_{L}^0, {(\nu_R^0)}^c)$ a column vector.
The matrix $\cal M $ is given by:
\begin{eqnarray}
{\cal M}= \left(\begin{array}{cc}  
 0  & m_D \\
m^T_D & M_R \end{array}\right) \label{calm}
\end{eqnarray}

It is always possible to choose a weak basis (WB) where the matrices 
$M_R$ and $m_l$ are both real and diagonal.   
The diagonalization of the $2n \times 2n$
matrix $\cal M$ 
is performed via the unitary transformation
\begin{equation}
V^T {\cal M}^* V = \cal D \label{dgm}
\end{equation}
where ${\cal D} ={\rm diag} (m_{\nu_1}, m_{\nu_2}, m_{\nu_3},
M_{\nu_1}, M_{\nu_2}, M_{\nu_3})$,
with $m_{\nu_i}$ and $M_{\nu_i}$ denoting the physical
masses of the light and heavy Majorana neutrinos, respectively. 
By writing $V$ and $\cal D$ in the following block form:
\begin{eqnarray}
V&=&\left(\begin{array}{cc}
K & R \\
S & T \end{array}\right) ; \label{matv}\\
{\cal D}&=&\left(\begin{array}{cc}
d & 0 \\
0 & D \end{array}\right)  \label{matd}
\end{eqnarray}
the leptonic charged current
interactions can be written in terms of mass eigenstates as:
\begin{equation}
{\cal L}_W = - \frac{g}{\sqrt{2}} \left( \overline{l_{iL}} 
\gamma_{\mu} K_{ij} {\nu_j}_L +
\overline{l_{iL}} \gamma_{\mu} R_{ij} {N_j}_L \right) W^{\mu}+h.c.
\label{phys}
\end{equation}
where $\nu_j$ and $N_j$ denote the light and heavy neutrinos.
Since the right-handed neutrino Majorana mass term is
SU(2) $\times$ U(1) invariant, the scale $M$ of $M_R$ can be much larger
than the scale  $v$ of electroweak symmetry breaking. Assuming 
$M^2 \gg v^2$ 
the light neutrino masses are obtained to an excellent approximation
from:
\begin{equation}
U_{\nu}^\dagger m_{eff}  U_{\nu}^* =d \label{14}
\end{equation}
where $m_{eff} = - m_D {M_R}^{-1} m^T_D$. 
The natural suppression of the eigenvalues of  $m_{eff}$ 
is the crucial point of the seesaw mechanism.
The unitary matrix $U_{\nu}$
obtained from Eq.~(\ref{14}) is the so-called Pontecorvo, Maki, Nakagawa
and Sakata matrix \cite{pmns} and coincides with $K$ 
in Eqs.~(\ref{matv}) and (\ref{phys}) up to corrections of order 
$\frac{v^2}{M^2}$, which we shall ignore under the above assumption.

In the WB where $m_l$, $M_R$ are diagonal, all mixing and
CP violation are contained in $m_D$ which is a complex $3 \times 3$
matrix. Three of its nine arbitrary 
phases can be eliminated by the simultaneous rephasing
of  $\nu_l$, $l_L$, so one is left with six CP violating
phases. Therefore, the three eigenvalues of $M_R$, together with
the 15 parameters of $m_D$ give a total of eighteen parameters.
This is to be compared with the nine parameters contained in the
low energy data, namely the three mixing angles and three
CP violating phases contained in $U_{\nu}$, together with the three
light neutrino masses.

The fact that there are many more parameters in $m_D$, $M_R$
than measurable quantities at low energies makes it impossible,
in general, to derive the seesaw parameters from low 
energy data. A particularly fascinating question is whether
it is possible to relate the size
and sign \cite{Frampton:2002qc}  of the observed
baryon asymmetry of the Universe (BAU) to  low energy CP violation,
in a framework of baryogenesis through leptogenesis. It has
been pointed out that this is only possible if further 
assumptions are introduced.
This can be readily seen by noting that, from Eq.~(\ref{14}) 
and the definition of $m_{eff}$, the matrix $m_D$ can be
parametrized as \cite{Casas:2001sr}:
\begin{equation}
m_D = i U_{\nu} {\sqrt d} G {\sqrt D_R} 
\label{udr}
\end{equation}
with $G$ an orthogonal complex matrix, 
${\sqrt D_R }$ a diagonal real
matrix verifying the relation 
${\sqrt D_R } {\sqrt D_R }= D_R $ 
and  ${\sqrt d }$ a real matrix with a maximum number
of zeros such that
${\sqrt d} {\sqrt d}^T = d $. Note that  ${\sqrt d }$
is not always a square matrix, as can be seen in section 3.
From Eq.~(\ref{udr}), it follows that:
\begin{equation}
m^\dagger_D m_D = -{\sqrt D_R} G^{\dagger}{\sqrt d}^T {\sqrt d} G  
{\sqrt D_R}  \label{drr}
\end{equation}
Since the CP violating phases relevant for leptogenesis 
are those contained in $m^\dagger_D m_D$ \cite{sym}, it is
clear that leptogenesis can occur even if there is
no CP violation at low energies i.e. no Majorana- or Dirac- type
CP phases at low energies \cite{Rebelo:2002wj}.

In the literature there have been various attempts at reducing the 
number of seesaw parameters by considering so-called minimal
scenarios. Models with only two right-handed neutrinos immediatly
lead to one massless light neutrino, whereas models with only one
right-handed neutrino would require two of the light
neutrinos to be massless and are, therefore, ruled out
in the context of type I seesaw, where no Higgs triplets are added.

\section{Example with one texture zero and
two right-handed neutrinos}
In this section, we 
reexamine the connection between leptonic low energy physics and 
leptogenesis, in the case of one texture zero and two 
right-handed neutrinos \cite{Ibarra:2003up}.
From the definition of $m_{eff}$, it can be readily realized that in the
case of only two right-handed neutrinos, one of the light neutrinos
is massless and one has:
\begin{equation}
U_{\nu}^\dagger m_{eff}  U_{\nu}^* = 
\left(\begin{array}{ccc}
0 &  &  \\
  & m_2 & \\
  &   & m_3  
\end{array}\right) 
\label{012}
\end{equation}
Let us assume now that $m_D$ has one texture zero \cite{exemplo}
and write it in the form:
\begin{equation}
m_D = \left(\begin{array}{cc}
a_1 & 0  \\
b_1  & b_2  \\
c_1  & c_2  
\end{array}\right) 
\label{32}
\end{equation}
with arbitrary non-zero entries, in the WB where $m_l$ and $M_R$ 
are diagonal and real. 
Let us write $m_D$ as in Eq.~(\ref{udr}), taking into account
that in this case, the complex orthogonal matrix $G$ is two-by-two
and can be parametrized by:
\begin{equation}
G = 
\left(\begin{array}{cc}
\cos Z  & \pm \sin Z   \\
- \sin Z  & \pm \cos Z   
\end{array}\right) 
\label{ggg}
\end{equation}
with Z complex. We use the following parametrization \cite{Eidelman:2004wy}
for $U_\nu$: 
\begin{eqnarray}
U_\nu =\left(
\begin{array}{ccc}
c_{12} c_{13} & s_{12} c_{13} & s_{13}  \\
-s_{12} c_{23} - c_{12} s_{23} s_{13}  e^{i \delta}
& \quad c_{12} c_{23}  - s_{12} s_{23} s_{13} e^{i \delta} \quad
& s_{23} c_{13}  e^{i \delta} \\
s_{12} s_{23} - c_{12} c_{23} s_{13} e^{i \delta}
& -c_{12} s_{23} - s_{12} c_{23} s_{13} e^{i \delta}
& c_{23} c_{13}  e^{i \delta}
\end{array}\right)\, \cdot P \ ,
\label{std}
\end{eqnarray}
where $c_{ij} \equiv \cos \theta_{ij}\ , \ s_{ij} \equiv \sin \theta_{ij}\ $
and $P={\rm diag\ }(1, e^{i \alpha /2}, 1)$; $\delta$ is a Dirac-type
phase and $\alpha$ is a
physical phase associated to the Majorana character of neutrinos.
In the general case, with three non-zero neutrino masses, 
two Majorana phases would be present.

In the case of only two right-handed neutrinos one has:
\begin{eqnarray}
d = \left(\begin{array}{ccc}
0 & 0 & 0 \\
0  & m_2 & 0 \\
0  &  0  & m_3  
\end{array}\right) \qquad 
\sqrt{d}  = \left(\begin{array}{cc}
0 & 0  \\
\sqrt{m_2}  & 0  \\
0  & \sqrt{m_3}  
\end{array}\right) \ . 
\label{drd}
\end{eqnarray}
From Eq.~(\ref{udr}) we then obtain:
\begin{eqnarray}
{m_D}_{i1} = i {U_\nu}_{i2} \sqrt{m_2} (\cos Z)  \sqrt{M_1} +
  i {U_\nu}_{i3} \sqrt{m_3} (-\sin Z)  \sqrt{M_1}  \\
{m_D}_{i2} = i {U_\nu}_{i2} \sqrt{m_2} (\pm \sin Z)  \sqrt{M_2} +
  i {U_\nu}_{i3} \sqrt{m_3} (\pm \cos Z)  \sqrt{M_2}  
\end{eqnarray}
The zero entry in $m_D$ implies:
\begin{equation}
s_{12} c_{13} e^{i \frac{\alpha}{2}}\  \sqrt{m_2} \ (\pm \sin Z) 
+ s_{13} \ \sqrt{m_3}\ (\pm \cos Z) =0
\label{zero}
\end{equation}
so that:
\begin{eqnarray}
\cos ^2 Z  =\frac{s_{12}^2 \  c_{13}^2 \ \rho}{ s_{12}^2 \  c_{13}^2 \ \rho +
s_{13}^2 \  e^{-i \alpha}} \ ;  \qquad
\frac{\sin Z}{\cos Z} = - \frac{ s_{13}e^{-i \frac{\alpha}{2}}}
{s_{12} c_{13} \sqrt{\rho}}  
\label{scg}
\end{eqnarray}
where $\rho = {m_2}/{m_3} $. Thus the zero texture in the matrix
$m_D$ allows for a full determination the matrix $G$, up to a reflexion, 
in terms of low energy measurable quantities. From Eq.~(\ref{drr}),
it is clear that knowledge of $G$ enables one to obtain the
phases appearing in $m^\dagger_D m_D$ which are the ones relevant 
for leptogenesis. Therefore, the presence of the texture zero 
leads to a connection between leptogenesis and low energy 
measurable quantities.

It is instructive to consider the case of a degenerate $M_R$.  
The expression for $m^\dagger_D m_D$, obtained from Eq.~(\ref{drr}),
becomes particularly simple:
\begin{equation}
m^\dagger_D m_D = \cal{N} \left(\begin{array}{cr}
s_{12}^2 \  c_{13}^2 \ \rho^2  + s_{13}^2 &  \qquad
\pm \sqrt{\rho} \ s_{12} \ c_{13}\ s_{13} (\rho\  e^{-i \alpha /2} -
 e^{i \alpha /2} )\\
h.c.  & \qquad s_{13}^2 \ \rho \ + \ s_{12}^2 \ c_{13}^2 \ \rho
\end{array}\right)
\label{exact}
\end{equation}
with $\cal{N} $ given by:
\begin{equation}
{\cal N} = M\ m_3 \ \frac{1}{(s_{12} \ c_{13}\ \sqrt{\rho})^2 \ 
\sqrt{|(1+ \lambda^2)|^2 }} \qquad 
\mbox{with} \  \lambda = \frac{ s_{13}}{s_{12} \ c_{13}\ \sqrt{\rho}} \ 
e^{-i \alpha /2}
\label{cal}
\end{equation}
where $M$ denotes the common heavy neutrino mass.
This simple example illustrates again 
the r\^ ole of texture zeros
in enabling to establish a connection between leptonic low energy physics 
and leptogenesis. Of course, exact degeneracy would have to be 
lifted in order for leptogenesis to be possible.
Almost degeneracy among heavy neutrinos leads to the
very interesting scenario of resonant leptogenesis \cite{resonance}.
Ibarra and Ross \cite{Ibarra:2003up} have analysed in detail the 
predictions from models with one and two texture zeros in $m_D$ in 
the case of two right-handed neutrinos, including the constraints
on leptogenesis and lepton flavour violating processes.
As pointed out in Ref. \cite{Ibarra:2003up}, the case of only one 
texture zero has the special feature of
fixing the matrix $G$ without imposing any further
restriction on light neutrino masses and mixing. This is clear from
Eq.~(\ref{udr}) which shows that each zero on each column
of $m_D$ corresponds to an orthogonality condition between that
column of the matrix $G$ and the corresponding row
of the matrix  $U_{\nu }\sqrt{d}$:
\begin{equation}
(m_{D})_{ij}=0~:\qquad (U_{\nu })_{ik}\sqrt{d}_{kl}G_{lj}=0  
\label{orto}
\end{equation}
With one zero in $m_D$ this equation has always a solution
for any $U_{\nu}$ and $\sqrt d$ of the form
given in Eq.~(\ref{drd}), independently of the hierarchy
between $M_1$ and $M_2$. 
Obviously, the r\^ ole of texture zeros
is to introduce restrictions which lead to the decrease in
the number of independent seesaw parameters. In particular
texture zeros lead to a decrease in the number of independent
CP violating phases.

As we have previously emphasized,  texture zeros are WB
dependent, in the sense that a texture zero present in one basis
may no longer exist in another WB. One may wonder whether
it is possible to translate particular texture zeros into
restrictions on the seesaw parameters expressed in terms of
WB invariants. The fact that texture zeros lead in general to a
decrease in the number of CP violating phases, provides a hint
that CP-odd WB invariants may be useful for introducing
restrictions on the seesaw parameters. In the next section
we address this question.

\section{On the relation between CP-odd WB invariants and texture 
zeros}
We start by recalling how CP-odd WB invariants can be constructed
by studying the CP properties of the present minimal extension of the
SM, which leads to the leptonic mass terms of Eq.~(\ref{lep}).
The starting point consists of writing the most general
CP transformation which leaves invariant the gauge interactions.
It can be readily seen that the CP transformation is given by:
\begin{eqnarray}
{\rm CP} l^0_L ({\rm CP})^{\dagger}&=&U^\prime
\gamma^0  C~ \overline{l^0_L}^T;  \quad
{\rm CP} l^0_R({\rm CP})^{\dagger}=V^\prime \gamma^0  C~ \overline{l^0_R}^T
\nonumber \\
{\rm CP} \nu^0_L ({\rm CP})^{\dagger}&=&U^\prime \gamma^0 C~
\overline{\nu^0_L}^T; \quad
{\rm CP} \nu^0_R ({\rm CP})^{\dagger}=W^\prime \gamma^0  C~
\overline{\nu^0_R}^T \label{cp}  \\
{\rm CP} W^+_{\mu} ({\rm CP})^{\dagger} &=& - (-1)^{\delta_{0\mu}}  
W^-_{\mu} \nonumber
\end{eqnarray}
where $U^\prime$, $V^\prime$, $W^\prime$ are unitary matrices 
acting in flavour space. The inclusion of these matrices reflects the
fact that in a WB, gauge interactions do not distinguish the various 
flavours.
Invariance of the mass terms under the above CP transformation,
requires that the following relations have to be satisfied:
\begin{eqnarray}
W^{\prime T} M_R W^\prime &=&-M_R^*  \label{cpM} \nonumber\\
U^{\prime \dagger} m_D W^\prime &=& {m_D}^*  \label{cpm} \\
U^{\prime \dagger} m_l V^\prime &=& {m_l}^* \label{cpml} \nonumber
\end{eqnarray}
From Eqs.~(\ref{cpm}) one can derive \cite{Branco:2001pq}
various CP-odd WB invariants, which are constrained to vanish
if CP invariance holds, following the procedure first outlined
in Ref.~\cite{Bernabeu:1986fc}. This procedure has been widely 
applied in the literature \cite{many1}
to the study of CP violation in many different scenarios.
An example is the following condition:
\begin{equation}
Tr[m_{eff} m^{\dagger}_{eff}, h_l]^3 = 0 \label{trc}
\end{equation}
where $h_l =  m_l m^\dagger_l$. This condition
is satisfied in the limit of no CP violation of Dirac type,
at low energies.

CP invariance requires the vanishing
of certain WB invariants. In the minimal seesaw model which we 
are considering, with an equal number of left-handed 
and right-handed neutrinos, in general
the number of CP violating phases equals 
$n^2 -n$,
where $n$ denotes the number of lepton flavours.
In the presence of flavour symmetries leading to texture
zeros and/or relations among parameters, one may have a smaller
number of CP violating phases and some of the CP-odd 
WB invariants may automatically vanish.

Next we analyse the possible connection between
texture zeros and
the vanishing of certain CP-odd invariants, in the cases of two
and three right-handed neutrinos.

\subsection{Two right-handed neutrinos and two texture zeros}
In the case of two right-handed neutrinos all ans\" atze 
with two texture zeros in $m_D$ have been studied in 
\cite{Ibarra:2003up}. It can be readily verified that in all
two texture zero ans\" atze the following WB invariant
condition is satisfied:
\begin{equation}
I_1 \equiv \mbox{tr} \left[ m_D M^\dagger_R M_R m^\dagger_D, h_l 
\right] ^3 = 0
\label{inv}
\end{equation}
One may ask whether the converse is also true, i.e., whether the
imposition of the condition of Eq.~(\ref{inv})
on arbitrary complex leptonic mass matrices 
automatically leads to one
of the two zero anz\" atze classified in \cite{Ibarra:2003up}.
We show that Eq.~(\ref{inv}), together with a reasonable
assumption of no conspiracy among the parameters of $m_D$
and those of $M_R$, does require the matrix $m_D$ 
to have two texture zeros in the WB 
where both $M_R$ and $m_l$ are diagonal real. Note that the 
hypothesis of ``no conspiracy'' is quite natural, since $m_D$ and 
$M_R$ originate in different terms of the Lagrangian.

In order to fix the notation let us write:
\begin{equation}
\label{mh}
m_D\ M_R^{\dagger }M_R\ m_D^{\dagger }=\left[
\begin{array}{ccc}
r_1 & \alpha _1 & \alpha _2 \\
\alpha _1^{*} & r_2 & \alpha _3 \\
\alpha _2^{*} & \alpha _3^{*} & r_3
\end{array}
\right]
\end{equation}
where $r_i$ are real and $\alpha _i$ are complex elements, which
depend on the heavy right-handed neutrino masses and the matrix elements
of $m_D$. Writing:
\begin{equation}
\label{md3}m_D \equiv \left[
\begin{array}{cc}
a_1 & a_2 \\
b_1 & b_2 \\
c_1 & c_2
\end{array}
\right]
\end{equation}
we obtain for the $\alpha _i$ in Eq. (\ref{mh})
\begin{equation}
\label{alfas3}
\begin{array}{ccc}
\alpha _1=M_1^2\ a_1b_1^{*}+M_2^2\ a_2b_2^{*} &  & \alpha _2=M_1^2\
a_1c_1^{*}+M_2^2\ a_2c_2^{*} \\
&  &  \\
\alpha _3=M_1^2\ b_1c_1^{*}+M_2^2\ b_2c_2^{*} &  &
\end{array}
\end{equation}
The WB invariant $I_1$, calculated in the WB where $m_l$ is also
diagonal, is given by:
\begin{equation}
\label{inv2}I_1=6i\ (m_\tau ^2-m_\mu ^2)(m_\tau ^2-m_e^2)(m_\mu ^2-m_e^2)\
{\rm Im}[\alpha _1\alpha _2^{*}\alpha _3]
\end{equation}
Clearly, $I_1=0$, if and only if
one of the $\alpha _i$'s is equal to zero 
or else the $\alpha _i$'s have cyclic
phases in such a way that $\arg [\alpha _1\alpha _2^{*}\alpha _3]=0, \pi $.
If one adopts the above ``no conspiracy'' hypothesis,
it is clear that the solutions
where one of the $\alpha_i$'s vanishes , would require that each
one of the two zeros contributing to that $\alpha_i$
should vanish. It can then be readily verified hat solutions of
Eq.~(\ref{inv}) in which one of the $\alpha _i$'s vanishes,
correspond to textures with one zero in each column.

For example
the requirement $\alpha _1=0$ is verified in the case of the following 
four  $m_D$ textures:
\begin{equation}
\label{aabb}
\left[
\begin{array}{cc}
0 & 0 \\
b_1 & b_2 \\
c_1 & c_2
\end{array}
\right]; \quad
\left[
\begin{array}{cc}
0 & a_2 \\
b_1 & 0 \\
c_1 & c_2
\end{array}
\right] ;\quad \left[
\begin{array}{cc}
a_1 & 0 \\   
0 & b_2 \\ 
c_1 & c_2
\end{array}      
\right]; \qquad
\left[
\begin{array}{cc}
a_1 & a_2 \\
0  & 0 \\
c_1 & c_2
\end{array}
\right] \quad
\end{equation}
All other possible textures with one zero in each column correspond
to either $\alpha _2=0$ or $\alpha _3=0$.
We consider now the solutions of Eq.~(\ref{inv}) corresponding to 
cyclic $\alpha _i$'s. It can be readily verified that cyclic solutions
correspond to textures with two zeros in the same column. Indeed, 
textures with two zeros in the first
column eliminate the terms with $M_2^2$ whilst terms with
two zeros in the second one eliminate the terms with $M_1^2$. An
example is:
\begin{equation}
\label{col}
\left[
\begin{array}{cc}
0 & a_2 \\
0 & b_2 \\
c_1 & c_2
\end{array}
\right]; \ 
\alpha _1=M_2^2\ a_2b_2^{*};\   \alpha _2=
M_2^2\ a_2c_2^{*}; \ 
\alpha _3=M_2^2\ b_2c_2^{*} 
\end{equation}
which obviously leads to  $\arg [\alpha _1\alpha _2^{*}\alpha _3]=0$.

All the fifteen different textures with two zeros
in $m_D$ are thus obtained from the invariant condition $I_1 = 0$,
together with the ``no-conspiracy'' hypothesis.
The low energy predictions arising from all these textures
were analysed in Ref.~\cite{Ibarra:2003up}, where it was shown 
that only five of them are allowed by present experiment.
It was also pointed out in \cite{Ibarra:2003up} that in general
texture zeros in $m_D$ may appear in a weak basis where 
neither $M_R$ nor $m_l$ are diagonal.
Implications for low energy physics in the case of non-diagonal
$M_R$ were also discussed \cite{Ibarra:2003up}, under certain 
restrictive assumptions on 
$m_l$. At this stage, it is worth noting  
that there are other CP-odd WB invariants which vanish
for all the two zero textures considered above, even if they arise
in a basis where $M_R$ is not diagonal. An example is the
following WB invariant condition:
\begin{equation}
I^\prime \equiv \mbox{tr} \left[ m_D m^\dagger_D, h_l 
\right] ^3 =0
\label{wbi}
\end{equation}
which is verified for any texture with two zeros in $m_D$
in a WB where $m_l$ is diagonal, while $M_R$ is arbitrary.

It should be pointed out that
two zeros in $m_D$, in the WB where $M_R$ and $m_l$ are
diagonal, still allow for CP violation. In fact, with two zeros 
in $m_D$ one can only eliminate at most two independent CP
violating phases out of the three present in the general
case, with two right-handed neutrinos. 
As a result, not all CP-odd WB invariants 
vanish in the case of two texture zeros. An example of
a CP-odd invariant  \cite{Branco:2001pq} which does not
vanish is given by:
\begin{equation}
I^{\prime \prime} \equiv \mbox{Im tr} \left[(m^\dagger_D  m_D)
M^{*}_R M_R M^{*}_R (m^\dagger_D  m_D)^\ast M_R \right]  \ .
\label{imtq}
\end{equation}
In the WB where $M_R$ is diagonal, it can be written as
\begin{equation}
I^{\prime \prime} = M_1 M_2 (M_2^2 - M_1^2) \mbox{Im}h^2_{12}
\label{wbitq}
\end{equation}
where
\begin{equation}
h_{12}= (m^\dagger_D  m_D)_{12} = a_1^{*}a_2 +
b_1^{*}b_2 + c_1^{*}c_2
\label{h12}
\end{equation}
so that $\mbox{Im}h^2_{12}$ can differ from zero in the case
of two texture zeros. This invariant is sensitive to the
combination $m^\dagger_D  m_D$ which is relevant for leptogenesis.
Furthermore, two texture zeros in $m_D$ also allow for
CP violation of Dirac type in the leptonic charged weak currents.
This can be seen from the condition written in Eq.~(\ref{trc}).

\subsection{Three right-handed neutrinos and three texture zeros}

Let us now consider the case of three right-handed neutrinos and analyse the
conditions under which the invariant $I_1$ vanishes.

In this case $m_{D}$ is a three-by-three matrix which can be written as: 
\begin{equation}
m_{D}=\left[ 
\begin{array}{ccc}
a_{1} & a_{2} & a_{3} \\ 
b_{1} & b_{2} & b_{3} \\ 
c_{1} & c_{2} & c_{3}
\end{array}
\right]  \label{md9}
\end{equation}
The parameters $\alpha _{i}$ of Eq. (\ref{mh}) are now given by: 
\begin{equation}
\begin{array}{ccc}
\alpha _{1}=M_{1}^{2}a_{1}b_{1}^{*}+M_{2}^{2}\ a_{2}b_{2}^{*}+M_{3}^{2}\
a_{3}b_{3}^{*} &  & \alpha _{2}=M_{1}^{2}a_{1}c_{1}^{*}+M_{2}^{2}\
a_{2}c_{2}^{*}+M_{3}^{2}\ a_{3}c_{3}^{*} \\ 
&  &  \\ 
\alpha _{3}=M_{1}^{2}b_{1}c_{1}^{*}+M_{2}^{2}\ b_{2}c_{2}^{*}+M_{3}^{2}\
b_{3}c_{3}^{*} &  & 
\end{array}
\label{alfas}
\end{equation}
for $M_{R}$ diagonal. Equation (\ref{inv2}) remains valid in the WB where 
$m_{l}$ is also diagonal. There are, as before, two types of possible
solutions.
Solutions in which one of the $\alpha _{i}$'s is zero (irrespective of 
$M_{R} $) are all those corresponding to three zeros in 
$m_{D}$ - one in each
column leaving one row without zeros, as for example in:
\begin{equation}
\left[ 
\begin{array}{ccc}
0 & 0 & a_{3} \\ 
b_{1} & b_{2} & 0 \\ 
c_{1} & c_{2} & c_{3}
\end{array}
\right] \ \ (\alpha _{1}=0),\ \ \left[ 
\begin{array}{ccc}
0 & a_{2} & a_{3} \\ 
b_{1} & b_{2} & b_{3} \\ 
c_{1} & 0 & 0
\end{array}
\right] \ \ (\alpha _{2}=0),\ \ \left[ 
\begin{array}{ccc}
a_{1} & a_{2} & a_{3} \\ 
b_{1} & 0 & b_{3} \\ 
0 & c_{2} & 0
\end{array}
\right] \ \ (\alpha _{3}=0)\ \   \label{bla}
\end{equation}
Solutions with three zeros in the same row would lead to one 
vanishing $\alpha _{i}$, but they are physically unnacceptable
since they correspond to the decoupling of one generation at low
energies. 

Any one of the $m_{D}$ matrices with three zeros has six real independent
parameters and three independent CP violating phases. Furthermore, we have
three Majorana masses $M_{1}$, $M_{2}$, $M_{3}$. This is to be compared to
three light neutrino masses, three mixing angles and three physical CP
violating phases at low energies. 

In addition to the solutions in Eq. (\ref{bla}), we obtain a set of cylic
solutions, similar to the case of two right-handed neutrinos in 
Eq. (\ref{col}), such that 
$\arg [\alpha _{1}\alpha _{2}^{\ast }\alpha _{3}]=0$ but
with $\alpha _{i}\neq 0$. However, these solutions correspond to four
texture zeros and therefore will be discussed in the 
following subsection, which is dedicated to the study of 
the connection between low and high energy CP violation in the 
context of models with four texture zeros.

\subsection{Three right-handed neutrinos and four texture zeros}

Cyclic solutions of Eq.~(\ref{inv}), in the case of three right-handed neutrinos, 
require, for arbitrary phases in $m_D$,
four zeros in this matrix.
In this case one column has no zeros, the other two columns have two zeros each,
as for example in:
\begin{equation}
\left[ 
\begin{array}{ccc}
0 & 0 & a_{3} \\ 
b_{1} & 0 & b_{3} \\ 
0 & c_{2} & c_{3}
\end{array}
\right] ,\left[ 
\begin{array}{ccc}
0 & a_{2} & 0 \\ 
0 & b_{2} & b_{3} \\ 
c_{1} & c_{2} & 0
\end{array}
\right]  \label{cyclic}
\end{equation}
Cyclic solutions where all zeros are grouped in one square, i.e.,
one column and one row have no zeros,  
are physically unacceptable, as they lead to $U_{i1}=0$ for some $i$.

For cyclic solutions of Eq.~(\ref{inv}), thus having four texture zeros, 
the number of parameters
in $m_{D}$ is much reduced. One has five real parameters and two complex
phases in $m_{D}$. In particular, for the non-squared cylic solutions in Eq.
(\ref{cyclic}), one may easily find the connection between leptogenesis
and low energy physics. 
Let us consider as an example, the first matrix in
Eq.~(\ref{cyclic}).  
The $a_i$, $b_i$ and $c_i$ in $m_{D}$ can be expressed as
functions of the neutrino masses, mixing angles and CP violating phases
through Eq.~(\ref{udr}). In this example, the matrix G can be 
fully determined by Eq.~(\ref{orto}) due to the existence of four zeros. 
With three right-handed neutrinos the matrix  ${\sqrt d }$ is diagonal
with nonzero entries $\sqrt {m_i}$ and we have, e.g.:
\begin{eqnarray}
(m_{D})_{12}=0~:\qquad (U_{\nu })_{1k}\sqrt{{m}_{k}}G_{k2}=0  \\
\label{calc}
\end{eqnarray} 
leading to:
\begin{eqnarray}
\left( \vec{G_1} \right)_i = \left( \varepsilon_{ijk} (U_{\nu })_{1j}
\sqrt{{m}_{j}} \; 
(U_{\nu })_{3k}\sqrt{{m}_{k}} \right) \; \frac{1}{N_1} \\
\left( \vec{G_2} \right)_i = \left( \varepsilon_{ijk} (U_{\nu })_{1j}
\sqrt{{m}_{j}} \; 
(U_{\nu })_{2k}\sqrt{{m}_{k}} \right) \; \frac{1}{N_2} \\
\left( \vec{G_3} \right)_i = \varepsilon_{ijk}
\left( \vec{G_1} \right)_j \;  \left( \vec{G_2} \right)_k 
= \frac{1}{N_3} \; (U_{\nu })_{1i}\sqrt{{m}_{i}} \qquad  \mbox{(no sum in i)}
\label{gggg}
\end{eqnarray} 
where the $\vec{G_i}$ are the columns of the matrix $G$ 
and the $N_{i}$ are complex normalization factors, with
phases such that $\vec{G_i}^2 = 1$.

Let us now consider the non-zero entries of $m_D$, for example $b_1$,
which corresponds to:
\begin{eqnarray}
(m_{D})_{21} =b_1=i \left( (U_{\nu })_{2k}\sqrt{{m}_{k}}\; \right) \; G_{k1} 
\sqrt{M_1} 
\label{only}
\end{eqnarray} 
Once the $G_{kj}$ are replaced by the explicit formulas obtained above, 
the coefficients of $m_D$ can be fully expressed in terms of physical quantities
only, up to non-physical phases which can be rotated away.

In this example we have
\begin{equation}
-m_{eff} = m_{D}\frac{1}{D} m_{D}^T=\left[ 
\begin{array}{ccc}
a_{3}^2 M_{3}^{-1} & a_{3}b_{3} M_{3}^{-1} & a_{3}c_{3} M_{3}^{-1} \\ 
a_{3}b_{3} M_{3}^{-1} & b_{1}^2 M_{1}^{-1} + b_{3}^2 M_{3}^{-1} & 
b_{3}c_{3} M_{3}^{-1} \\ 
a_{3}c_{3} M_{3}^{-1} & b_{3}c_{3} M_{3}^{-1} & 
c_{2}^2 M_{2}^{-1} +c_{3}^2 M_{3}^{-1} 
\end{array}
\right]  
\label{abc}
\end{equation}
Only five entries in $m_{eff}$ are independent. We can relate 
$(m_{eff})_{23}$ to other entries by:
\begin{equation}
(m_{eff})_{11} (m_{eff})_{23} - (m_{eff})_{12} \; (m_{eff})_{13} =0
\label{mmmm}
\end{equation}
This relation implies low energy constraints and 
furthermore guarantees the orthogonality of columns 
one and two of the matrix $G$, as defined above.
It is clear from the definition of $m_{eff}$ that $G$ does
not play any r\^ ole in low energy physics. However in the matrix
$m_{D}^{\dagger }m_{D}$, which is the matrix relevant for 
leptogenesis, the elements of $G$ play an important r\^ ole
since they do not cancel out.
In this example, there is a strong relation between leptogenesis and low energy
physics due to the fact that G can be fully expressed in terms of
measurable low energy parameters. With four texture zeros, there are
constraints in the low energy physics which result from the reduction of
the independent parameters in $m_{eff}$  as expressed 
in this example by Eq.~(\ref{mmmm}). This relation excludes
scenarios with direct or inverse hierarchical light neutinos, i.e.
the case of one neutrino mass much smaller than the other two. 
Likewise, in the three zero textures of section 4.2, there 
are also low energy constraints which in this case translate into 
the existence of one zero in one of the off-diagonal elements
of $m_{eff}$ (and its symmetric entry). For instance, for the
first  matrix in Eq. (\ref{bla}), corresponding to the case 
$\alpha _{1}=0$, one has $(m_{\mathrm{eff}})_{12}=0$, or
equivalently
\begin{equation}
m_1 U_{11}U_{21}+m_2 U_{12}U_{22}+m_3 U_{13}U_{23}=0
\label{m12zero}
\end{equation}
In Ref.\cite{Hagedorn:2004ba} the stability
of zeros in neutrino mass matrices under quantum corrections, in
type I seesaw models, has been studied. It was found that
some of the two-zero textures for the neutrino mass matrix
that have been classified as incompatible with experimental data, are
not excluded. 
A detailed study of the phenomenology of three and four texture zeros
in $m_D$ is beyond the scope of this paper.

Four texture zeros may also be obtained from the solutions with three
texture zeros  considered in the previous subsection in 
which one of the $\alpha _{i}$'s is zero. 
However, for these cases, the extra zero has to be imposed by
demanding that a new invariant $I_{2}$ vanishes\footnote{
With respect to the cyclic solutions, we do not need to consider this
invariant, as they automatically obey $I_{2}=0$.} . Taking 
\begin{equation}
I_{2}\equiv \mathrm{ tr \ }\left[ M_{R}^{\dagger }M_{R}\ ,\ \ m_{D}^{\dagger
}m_{D}\right]^3   \label{i2}
\end{equation}
and computing $I_{2}$ for e.g. the $\alpha _{1}=0$ case in Eq. (\ref{bla}),
one finds 
\begin{equation}
I_{2}=6i\ (M_{3}^{2}-M_{2}^{2})(M_{3}^{2}-M_{1}^{2})(M_{2}^{2}-M_{1}^{2})\
|c_{3}|^{2}\ \mathrm{Im}[b_{1}^{\ast }b_{2}c_{1}c_{2}^{\ast }]  \label{i2a}
\end{equation}

It is clear that $I_{2}=0$ , if one of the parameters\footnote{
Cases where e.g. $\arg [b_{1}^{*}b_{2}c_{1}c_{2}^{*}]=0$, or with other phase 
relations amongst the $a$'s, $b$'s and $c$'s will not be studied
here.} $b_{1}$, $b_{2}$, $c_{1}$, $c_{2}$ vanishes. The case $c_{3}=0$ is of no
physical interest as it leads to vanishing solar neutrino mixing, which is
clear by computing $m_{\mathrm{eff}}$. Taking e.g. $c_{1}=0$, one then
obtains for $m_{D}$ 
\begin{equation}
m_{D}=\left[ 
\begin{array}{ccc}
0 & 0 & a_{3} \\ 
b_{1} & b_{2} & 0 \\ 
0 & c_{2} & c_{3}
\end{array}
\right]   \label{mdd1}
\end{equation}
which has 4 texture zeros. 

It is interesting to note that 
imposing $I_{2}$ equal to zero, irrespective of condition
$I_1 =0$, for non degenerate $M_i$, requires:
\begin{equation}
{\rm Im} [(m_{D}^{\dagger}m_{D})_{12}(m_{D}^{\dagger}m_{D})_{31}
(m_{D}^{\dagger}m_{D})_{23}] =0
\label{mais}
\end{equation}
where
\begin{eqnarray}
(m_{D}^{\dagger}m_{D})_{12} = a_1^* a_2 +b_1^* b_2 + c_1^* c_2 
\nonumber \\ 
(m_{D}^{\dagger}m_{D})_{13} = a_1^* a_3 +b_1^* b_3 + c_1^* c_3 \\ 
(m_{D}^{\dagger}m_{D})_{23} = a_2^* a_3 +b_2^* b_3 + c_2^* c_3 
\end{eqnarray}
Matrices $m_D$, with 
three zeros, one on each row leaving
one column without zeros, such as:
\begin{eqnarray}
\left[ 
\begin{array}{ccc}
0 & a_{2} & a_{3} \\ 
0 & b_{2} & b_{3} \\ 
c_{1} & 0 & c_{3}
\end{array}
\right] ,  
\left[ 
\begin{array}{ccc}
a_{1} & a_{2} & 0 \\ 
0 & b_{2} & b_{3} \\ 
0 & c_{2} & c_{3}
\end{array}
\right] ,
\left[ 
\begin{array}{ccc}
a_{1} & 0 & a_{3} \\ 
b_{1} & 0 & b_{3} \\ 
c_{1} & c_{2} & 0
\end{array}
\right]\; .
\label{blabla}
\end{eqnarray}
verify this condition. These matrices are the transposed of the
solutions found in section 4.2.

\section{Conclusions}
We have shown that CP-odd WB invariants can be useful
in the analysis of lepton flavour models with texture zeros.
In particular, we have pointed out that there is a large
class of sets of texture zeros considered in the literature
which lead to the vanishing of certain CP-odd invariants.
Conversely, it was shown that starting from arbitrary complex
leptonic mass matrices, the imposition of the vanishing
of certain CP-odd invariants together with a reasonable
assumption of no conspiracy among the parameters of $m_D$ and $M_R$,
automatically leads to given sets of texture zeros. 
These WB invariants enable one to recognize models characterized 
by texture zeros in $m_D$ in the WB where $m_l$, $M_R$ are diagonal, 
when these same models are written in a different WB where the
texture zeros are not manifest.

We have also discussed the r\^ ole of texture zeros in allowing
for a connection between leptogenesis and low energy data, such as 
leptonic masses, mixing and CP violation. We have done the analysis
in the context of two, three and four texture zeros. The crucial
point is the fact that in the presence of texture zeros, the matrix
$G$ defined in Eq.~(\ref{udr}), can be expressed in terms of low 
energy parameters. Recall that $G$ enters in ${m_D}^\dag m_D$
which in turn plays a crucial r\^ ole in leptogenesis. Furthermore 
texture zeros lead in general to specific predictions at low
energies. 

An important step towards the understanding of the flavour puzzle
would be finding a theoretical framework which would naturally lead
to the vanishing of the CP-odd invariants considered in this
paper or else to specific texture zeros.

\section*{Acknowledgements}
This work was partially supported by CERN and by Funda\c{c}\~{a}o para a 
Ci\^{e}ncia e a  Tecnologia (FCT, Portugal), through the projects 
POCTI/FNU/44409/2002, PDCT/FP/
FNU/50250/2003,
POCI/FP/63415/2005, POCTI/FP/FNU/50167/2003
and CFTP-FCT UNIT 777, which
are partially funded through POCTI (FEDER). The authors are grateful
for the warm hospitality of the CERN Physics Department (PH) Theory (TH)
where this work started. The work of G.C.B. was supported by the Alexander 
von Humboldt Foundation through a Humboldt Research Award. G.C.B. would
like to thank Andrzej J. Buras for the kind hospitality at TUM and M.N.R. 
thanks Wolfgang Hollik for the warm hospitality at
the Max-Planck--Institut f\" ur Physik (Werner--Heisenberg--Institut)
and also for support.

\end{document}